\documentclass[showpacs,preprintnumbers,amsmath,amssymb,prb,twocolumn]{revtex4}

\usepackage{graphicx}
\usepackage{dcolumn}
\usepackage{amsmath}
\usepackage{amssymb}
\usepackage{bm}
\usepackage{xspace}

\newcommand{\deltaT}{\Delta T}
\newcommand{\Int}{\int}
\newcommand{\IntL}[2]{\int_{#1}^{#2}}
\renewcommand{\d}[1]{d#1}

\newcommand{\Tr}{\mathsf{Tr}}
\newcommand{\vdist}{\mathsf{d}}

\newcommand{\hscale}{0.17}
\newcommand{\vscale}{0.19}

\newcommand{\metal}[1]{#1}
\newcommand{\IA}{IA\xspace}
\newcommand{\IIA}{IIA\xspace}
\newcommand{\IB}{IB\xspace}
\newcommand{\IIB}{IIB\xspace}
\newcommand{\charge}{q}

\newcommand{\Eqref}[1]{Eq.~\ref{#1}}
\newcommand{\DoubleEqref}[2]{\Eqref{#1} and \Eqref{#2}}
\newcommand{\Figref}[1]{Fig.~\ref{#1}}
\newcommand{\DoubleFigref}[2]{Figs.~\ref{#1} and \ref{#2}}
\newcommand{\Citeref}[1]{Ref.~\onlinecite{#1}}
\newcommand{\secref}[1]{section \ref{#1}}
 
\newcommand{\bvec}[1]{\mathrm{\mathbf{#1}}}
\newcommand{\typeI}{type-I\xspace}
\newcommand{\typeII}{type-II\xspace}

\newcommand{\xIII}{x_{\mathrm{I/II}}}
\newcommand{\kIII}{\kappa_{\mathrm{I/II}}}
\newcommand{\xct}{x_{\text{tri}}}
\newcommand{\kct}{\kappa_{\text{tri}}}

\begin{document}

\title{Vortex Interactions and Thermally Induced Crossover from Type-I to Type-II Superconductivity}
\author{J. Hove}
  \email{Joakim.Hove@phys.ntnu.no}
  
\author{S. Mo} 
  \email{Sjur.Mo@phys.ntnu.no}
  
\author{A. Sudb{\o}}
  \email{Asle.Sudbo@phys.ntnu.no}
  
  \affiliation{%
     Department of Physics \\
     Norwegian University of Science and Technology, \\
     N-7491 Trondheim, Norway}

\date{\today}



\begin{abstract}
  We have computed the effective interaction between vortices in the 
  Ginzburg-Landau model from large-scale Monte-Carlo simulations, taking
  thermal fluctuations of  matter fields and gauge fields fully into account 
  close to the critical temperature. We find a change, in the form of what 
  appears to be a crossover, from an attractive to a repulsive 
  effective vortex interactions in an intermediate range of Ginzburg-Landau 
  parameters $\kappa \in [0.76,1]/\sqrt{2}$ upon increasing the temperature 
  in the superconducting state. This corresponds to a thermally induced 
  crossover from \typeI to \typeII superconductivity around a temperature 
  $T_{\rm{Cr}}(\kappa)$, which we map out in the vicinity of the 
  metal-to-superconductor transition. In order to see this crossover, it is 
  essential to include  amplitude fluctuations of the matter field, in 
  addition to phase-fluctuations and gauge-field fluctuations.
  We present a simple physical 
  picture of the crossover, and relate it to observations in \metal{Ta}
  and \metal{Nb} elemental superconductors which have low-temperature
  values of $\kappa$ in the relevant range.
\end{abstract}
\maketitle


\section{Introduction}
The nature of the phase-transition in systems of a scalar matter field
coupled to a massless gauge-field has a long history in condensed
matter physics, dating at least back to the introduction of the
Ginzburg-Landau (GL) theory of superconductivity \cite{Ginzburg:1950}.
At the mean-field level, ignoring spatial variations in gauge fields
as well as matter fields leads to the prediction of a second order
phase transition in the model, with classical mean-field exponents for
all values of the GL parameter $\kappa$. The first attempt to
seriously consider the role of fluctuations on the order of the
metal-to-superconductor transition was made by Halperin, Lubensky, and
Ma \cite{Halperin:1974}, who found that ignoring matter-field
fluctuations entirely, and treating gauge-field fluctuations exactly,
resulted in a permanent first order transition for all values of
$\kappa$, since the gauge-field-fluctuations produced an extra term
$\sim - |\phi|^3$ in the matter field sector of the theory in three
spatial dimensions, where the complex matter field is denoted by
$\phi$ and represents the condensate order parameter. (In the context
of particle physics, Coleman and Weinberg \cite{Coleman:1973} studied
the equivalent problem of spontaneous symmetry breaking due to
radiative corrections in the Abelian Higgs model in four space-time
dimensions, finding the additional term $\phi^4 \ln(\phi^2/\phi_0^2)$,
where the real matter field is denoted $\phi$ and represents a scalar
meson.) \cite{Note_order_from_disorder}. Subsequently, Dasgupta and
Halperin\cite{Dasgupta:1981} found, using duality arguments in
conjunction with Monte-Carlo simulations, that when gauge-field
fluctuations and phase fluctuations of the scalar matter field are
taken into account, but amplitude fluctuations are ignored, the phase
transition is permanently second order\cite{Dasgupta:1981}.
Bartholomew \cite{Bartholomew:1983} then reported results from
Monte-Carlo simulations for the case when also amplitude fluctuations
are taken into account, concluding that the phase transition changes
from first to second order at a particular value of the GL parameter
$\kappa \approx 0.4/\sqrt{2}$. As far as this numerical value is
concerned, note that the problem of finding a {\it tricritical } value
$\kappa_{\rm{tri}}$ separating first and second order transitions is
extremely demanding even by present day supercomputing standards
\cite{Mo:2001} (see below). Using ingenious duality arguments,
Kleinert \cite{Kleinert:1982} obtained that the change from first to
second order transition should occur at $\kappa \approx 0.8/\sqrt{2}$.
The value of $\kappa$ that separates a first order (discontinuous)
transition from a second order (continuous) one, defines a tricritical
point \cite{Note_tricritical}, and will hereafter be denoted $\kct$.
Note that to obtain the above result, it is necessary to allow
for amplitude fluctuations in the superconducting order parameter,
which become important for small to intermediate values of $\kappa$,
but are totally negligible in the extreme \typeII regime 
$\kappa >> 1$.

The critical properties of a superconductor may be investigated at the
phenomenological level by the GL model of a complex scalar matter
field $\phi$ coupled to a fluctuating massless gauge field $\bvec{A}$.
It is this feature of the gauge field that makes the GL model so
difficult to access by the standard techniques employing the
renormalisation group \cite{Halperin:1974,Note}. The GL model in $d$
spatial dimensions is defined by the functional integral
\begin{widetext}
\begin{align}
 Z &= \int \mathcal{D}A_\nu \mathcal{D}\phi 
\exp [- \int d^dx  \left[\frac{1}{4}  F_{\mu \nu}^2 + |(\partial_\nu +i \charge A_\nu) \phi|^2 
+ m^2 |\phi|^2 + \lambda |\phi|^4 \right] 
\label{GLM}
\end{align}   
\end{widetext}
where $F_{\mu \nu } = \partial_\mu A_\nu - \partial_\nu A_\mu$, $\charge$ is the 
charge coupling the condensate matter field $\phi$ to the fluctuating gauge field 
$A_\mu$, $\lambda$ is a self coupling, and $m^2$ is a mass parameter which changes 
sign at the mean field critical temperature. When all dimensionful quantities are 
expressed in powers of the scale represented by $\charge^2$, the GL model may be 
formulated in terms of the two dimensionless parameters $y= m^2/\charge^4$ and 
$x=\lambda/\charge^2$. In this case, $y$ is temperature like and drives the system 
through a phase transition, and $x= \kappa^2$ is the square of the Ginzburg-Landau  
parameter. Depending on the value of $x$, the transition is either first order for 
$x < \xct$, or continuous for $x > \xct$ \cite{Kleinert:1982,Bartholomew:1983,Mo:2001}.  

In a recent paper\cite{Mo:2001}, we have determined $\xct = 0.295 \pm
0.025$.  This corresponds to a tricritical value of the GL parameter
$\kct=(0.76 \pm 0.04)/\sqrt{2}$, in rather remarkable agreement with
the results of \Citeref{Kleinert:1982}. {\it Moreover in
  \Citeref{Mo:2001} it was also argued that this value of $x$ or
  $\kappa$ is the demarkation value which separates \typeI and \typeII
  superconductivity, rather than the classical mean-field value
  $\kappa = 1/\sqrt{2}$.} The connection can be made when one realizes that
{\it criticality} at the metal-to-superconductor transition requires
that topological defects of the matter field in the form of vortex
loops are stable.  On the other hand, there is a connection between
critical exponents and geometric properties of a tangle of such vortex
loops\cite{Hove:2000}. The fractal dimension $D_H$ of the vortex-loop
tangle is connected to the anomalous scaling dimension $\eta_\phi$
\cite{Nguyen:1999} of the matter field in a field theory of the
vortex-loop gas, a theory dual to the original GL theory
\cite{Tesanovic:1999,Nguyen:1999} by the relation $D_H + \eta_\phi=2$.
Since the anomalous scaling dimension is connected to the order
parameter exponent $\beta$ of the dual matter field by the relation $2
\beta = \nu (d-2 + \eta_{\phi})$ \cite{Nguyen:1999,Hove:2000}, it
follows that a collapse of the vortex-loop tangle implies $D_H=d$ and
hence $\beta=0$ indicative of a first order transition. Here, $d$ is
the spatial dimension of the system. Now, a collapse of the tangle in
turn implies an effective attraction between vortices, or \typeI
behavior. On the other hand, a stable vortex-loop tangle at the
critical point, with fractal dimension $D_H < d$, implies first of all
\typeII behavior, but also $\eta_\phi > 2-d$ and $\beta > 0$, and
hence a second order transition.

The above assertion, that the tricritical value of $\kappa$ separates
first order from second order metal-to-superconductor transition, and moreover
also separates \typeI from \typeII behavior when the system is on the
phase-transition line $y_c(x)$, is in  contrast to the conventional wisdom 
that \typeI and \typeII superconductivity is separated by $x=0.5$. Based on 
the above arguments, we have proposed the phase diagram shown in Fig. 1 of 
\Citeref{Mo:2001} which contains a new line separating \typeI and \typeII 
superconductivity. The shape of this line was inferred from the observation 
that far from the phase transition, mean-field estimates of the boundary between 
\typeI and \typeII should be precise, and hence this boundary should 
asymptotically approach $x=0.5$ from below as the temperature is reduced. 

It is the purpose of this paper to show directly, by computing the
effective thermally renormalized interaction between vortices via
large-scale Monte-Carlo simulations, that this quantity changes from
being repulsive to attractive in the intermediate regime $\kappa \in
[0.76,1]/\sqrt{2}$. Since the sign of the vortex-interaction is the
microscopic diagnostics, in terms of vortex degrees of freedom, for
distinguishing \typeI from \typeII superconductivity, the large-scale
simulations we present in this paper confirm the above conjectures and
plausibility arguments of \Citeref{Mo:2001}.

In an external field the GL model has classical solutions in terms of
Abrikosov flux tubes\cite{Abrikosov:1957b}, or Nielsen--Olesen
vortices\cite{Olesen:1973}, and the concept of \typeI versus \typeII
superconductivity is based on the interaction between these vortices.
For \typeI superconductors they attract each other, whereas for
\typeII superconductors the interaction is repulsive.
Abrikosov\cite{Abrikosov:1957b} showed that at the \emph{mean field}
level \typeI and \typeII superconductors are separated at $\kappa =
1/\sqrt{2}$.  We will refer to the value of $\kappa$ separating \typeI
from \typeII behavior at $\kIII$, which we find varies with $y$. It is
not a sharply defined quantity, since it represents a crossover line.
The exception is at $y=y_c,x=\xct$, where $\kIII=\kct$.  Elaborate
calculations of vortex interactions have been carried out
\cite{Hartmann:1966,Kramer:1971,Bogomolnyi:1976,Bogomolnyi:1976b,Jacobs:1979},
but none of these approaches take thermal fluctuations into account. A
recent overview of superconductors with $\kappa$ close to $1/\sqrt{2}$
can be found in Ref. \onlinecite{Lukyanchuk:2001}, see also
Ref. \onlinecite{KleinNog}.

Superconductors with $\kappa \approx 1/\sqrt{2}$ were
studied extensively in the 1960s and 1970s\cite{Lukyanchuk:2001}, and
in particular measurements on the metals \metal{Ta} and \metal{Nb}
demonstrated that the notion of a \emph{temperature independent} value of
$\kIII$ was incorrect\cite{Auer:1973}. At the time, this was explained
with a mean-field theory involving three GL parameters 
\cite{Maki:1969}. Thermal fluctuations, not addressed at 
the mean-field level, offer an alternative  and above all simpler 
explanation for the observations of crossovers from \typeI to \typeII 
behavior in one and the same compound as the temperature is increased. 

We have performed large scale Monte Carlo (MC) simulations on the
lattice version of \Eqref{GLM}, with two vortices penetrating the
sample in the $\hat{z}$ direction. By measuring the interaction
between these two vortices we have determined the value of $\kIII$,
in particular how this value is affected by thermal fluctuations close
to the critical point.

\begin{widetext}
\section{Model, simulations and results}
To perform simulations on \Eqref{GLM}, we have defined a discrete
version as follows\cite{Kajantie:1999}:
\begin{align}
  \label{GLML}
  Z &= \int \mathcal{D} \bm{\alpha} \mathcal{D}\psi \exp(-S[\bm{\alpha},\psi]) \notag \\
  S[\bm{\alpha},\psi] &= \beta_G \sum_{\bvec{x},i<j} \frac{1}{2} \alpha_{ij}(\bvec{x})^2 - 
  \frac{2}{\beta_G} \sum_{\bvec{x},\hat{\imath}} \mathsf{Re} \left[\psi^{\ast}(\bvec{x})e^{i\alpha_i(\bvec{x})} \psi(\bvec{x} + \hat{\imath})\right] + 
  \beta_2 \sum_{\bvec{x}} \psi^{\ast}(\bvec{x}) \psi(\bvec{x}) + 
  \frac{x}{\beta_G^3} \sum_{\bvec{x}} \left[ \psi^{\ast}(\bvec{x}) \psi(\bvec{x}) \right]^2.
\end{align}
In \Eqref{GLML} $\alpha_i(\bvec{x}) = a q A_i(\bvec{x})$ and
$\alpha_{ij}=\alpha_i(\bvec{x})+\alpha_j(\bvec{x}+\hat{\imath})
-\alpha_j(\bvec{x})-\alpha_j(\bvec{x}+\hat{\jmath})$. $\beta_G$ and
$\beta_2$ are related to the continuum parameters $x$ and $y$ and the
lattice constant $a$,
\begin{align}
  \label{}
  \beta_G &= \frac{1}{a \charge^2} \\
  \label{betaIIdef}
  \beta_2 &= \frac{1}{\beta_G} \bigg[6 + \frac{y}{\beta_G^2} - \frac{3.1759115\left(1 + 2x\right)}{2 \pi \beta_G}
            - \frac{\left(-4 + 8x - 8x^2\right) \left( \ln 6\beta_G + 0.09\right) - 1.1 + 4.6x}{16 \pi^2 \beta_G^2} \bigg].
\end{align}
\end{widetext}
Note that $\beta_2$ contains the effect of ultraviolet renormalization
in the continuum limit when the lattice constant $a \to
0$\cite{Laine:1998,Mo:2001}. The model \Eqref{GLML} is defined on a
numerical grid of size $N_x \times N_y \times N_z$, corresponding to a
\emph{physical} size of $L_x \times L_y \times L_z$, with $L_i = N_i
a$. All our simulations have been on cubic systems with $\beta_G = 1$.

To impose an external magnetic field\cite{Kajantie:1999}, we modify
the action \Eqref{GLML} by changing the field energy along one stack
of plaquettes located at $x_0,y_0$ in such a way that the action is
minimized for $\alpha_{12}(x_0,y_0,z) = -2\pi n$ instead of zero,
corresponding to forcing a number of $n$ flux quanta through the
system. Hence, the action $S\left[\alpha,\psi;n \right]$ for $n$ 
flux-quanta forced through the system, is given by 
\begin{widetext}
\begin{equation}
\label{fieldGL}
S\left[\alpha,\psi;n\right] 
= S\left[\alpha,\psi;0\right] + \sum_{z} \left(2\pi n \alpha_{12}(x_0,y_0,z) + 2\pi^2n^2\right).
 \end{equation}
\end{widetext}
The second term in \Eqref{fieldGL} corresponds to forcing a flux
$\Phi_B$ through the lattice in the negative $z$-direction
\begin{equation}
  \label{flux}
  \frac{\alpha_{12}(x_0,y_0,z)}{\charge} = a^2 \charge \left( \nabla \times \bvec{A}(x_0,y_0,z) \right)_z 
= -\frac{2\pi n}{\charge}.
\end{equation}

The crucial point is that, due to periodicity, the total flux
through the system \emph{must be zero}, i.e.
\begin{equation*}
  \sum_{x,y} \alpha_{12}(x,y,z) = 0 \quad \forall z. 
\end{equation*}
Consequently, the $n$ flux-quanta of the total flux $2\pi n/\charge$
must return in the $+z$ direction. This flux returns in a manner
specified by the dynamics of the theory,\cite{Kajantie:1999} and it is
this \emph{response} which is the topic of interest in the current
paper.

The experimental situation corresponds to applying an external magnetic 
field $H$, and then study the magnetic response of the superconductor to 
this field. Hence, a suitable thermodynamic description is coached in terms 
of a potential $\Phi(H)$, which is a function of the \emph{intensive} field
variable $H$. In the simulations we have fixed $n$, which is analogous
to fixing the magnetic induction, and a description based on the
\emph{extensive} field variable $B$ is more appropriate. The two
approaches are related by a Legendre transformation\cite{Kajantie:1999}.
In principle the simulations could also be performed in an ensemble
with fixed magnetic field. Technically this would be achieved by
adding the term
\begin{equation}
  \label{}
  HL_z 2\pi n/q \nonumber
\end{equation}
to the action in \Eqref{GLM}. This would promote $n$ to a dynamical
variable of the theory, and be more in accordance with the
experimental situation. However, a change $n \to n \pm 1$ would
require a global relaxation, and this would give very low acceptance
rates, i.e. inefficient simulations. 

For \typeI superconductors, superconductivity vanishes for $H > H_c$.
For \typeII superconductors, a \emph{flux line lattice} is formed
at $H = H_{c_1}$, for smaller fields the magnetization in the sample
vanishes due to the Meissner effect.  By fixing $n$ one can not study
these effect directly, however it is possible to determine a
corresponding field strength from $n$, see \Citeref{Kajantie:1999}.

On the basis of simulations performed using the modified action
\Eqref{fieldGL}, we have determined the {\it effective}
temperature-renormalized interaction between two vortex-lines, and
searched for the value of the GL parameter, or more precisely its
square, $\xIII$, where this interaction changes character from being
effectively attractive to being effectively repulsive. In
\secref{section.two} we have fixed $n=2$ and studied the distance
between the flux lines. In \secref{section:free.energy} we have
generalized to real $n$, and used this to calculate a free energy
difference between states containing one and two vortices. This is
also a measure of the sign of the effective vortex interaction, and
hence an indication of whether we have \typeI or \typeII behavior.

To obtain the results in Figs. \ref{FSS}, \ref{VFig}, and
\ref{DistFig} we have performed simulations on cubic systems of size $N =
8,12,16,24,32,48$, with $\beta_G = 1.00$. All simulations have been 
performed in the broken symmetry state $y < y_c$, with particular emphasis 
on the values $y = -0.04, -0.10, -0.20, -0.30, -0.40$. For the two largest
system sizes the final datapoints are averages of approximately $10^6$
sweeps, whereas approximately $10^5$ sweeps have been performed for
the four smallest system sizes. 

The simulations leading to the results
of \Figref{deltaTFig} are quite different. They are performed for the
fixed system parameters $N=24$ and $\beta_G = 1.00$, and for
each value of $m$, we have performed from $2.5 \cdot 10^4$ to 
$2.5 \cdot 10^5$ MC sweeps through the lattice.  One sweep
through the lattice consists of (1) conventional local Metropolis
updates of $\psi$ and $\bvec{A}$, and (2) global radial updates of
$\left| \psi\right|$ combined with  overrelaxation
\cite{Kajantie:1996,Dimopoulos:2000} of $\psi$.

\subsection{Effective vortex interaction}
\label{section.two}
We first clarify what is meant by {\it effective vortex interaction} 
in this context. In the Ginzburg-Landau model at zero temperature,
one may compute a pair-potential between two vortices which
consists of an attractive part due to vortex-core overlap,
and a repulsive part due to circulation of supercurrents (or
magnetic fields) outside the vortex core. Ignoring fluctuation 
effects, this furnishes an adequate way of distinguishing between 
\typeI and \typeII behavior, by asking when the attractive 
core-contribution dominates the magnetic field contribution or 
vice versa.  By effective interaction, we mean a thermally renormalized
pair interaction which fully takes entropic contributions into
account. At low temperatures the effective interaction will
revert back to the standard pair-interaction described above, but will
deviate as temperature is raised, and this is particularly relevant
as the critical temperature is approached, as we shall see below.
We also comment further on this in the Discussion  section,
where we elaborate on what we perceive to be a crossover 
between \typeI and \typeII behavior. 

\begin{figure}[htbp]
\begin{center}
\begin{tabular}{cc}
  \scalebox{\hscale}[\vscale]{\rotatebox{-90.0}{\includegraphics{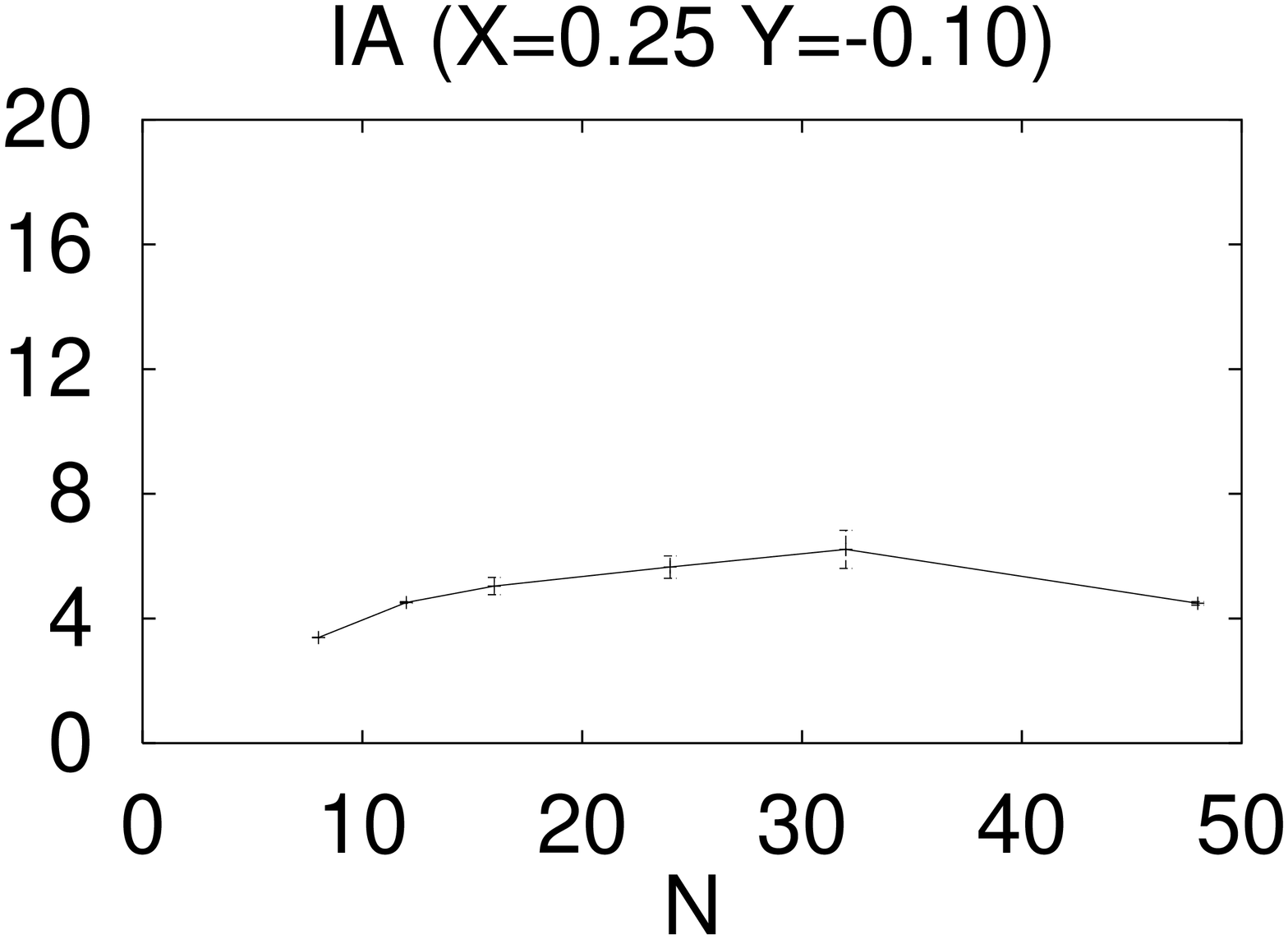}}} & 
  \scalebox{\hscale}[\vscale]{\rotatebox{-90.0}{\includegraphics{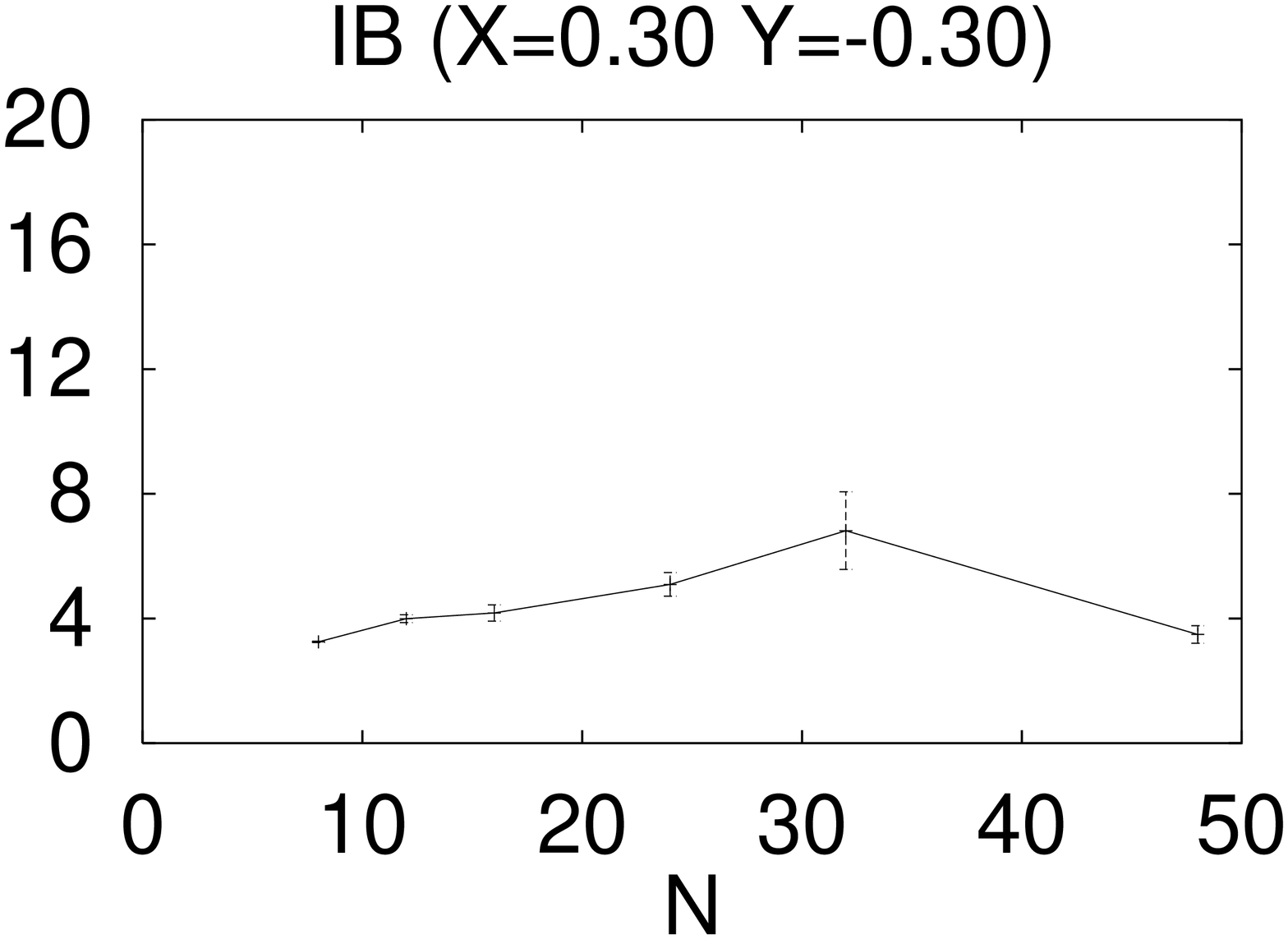}}}
  \\                                                             
  \scalebox{\hscale}[\vscale]{\rotatebox{-90.0}{\includegraphics{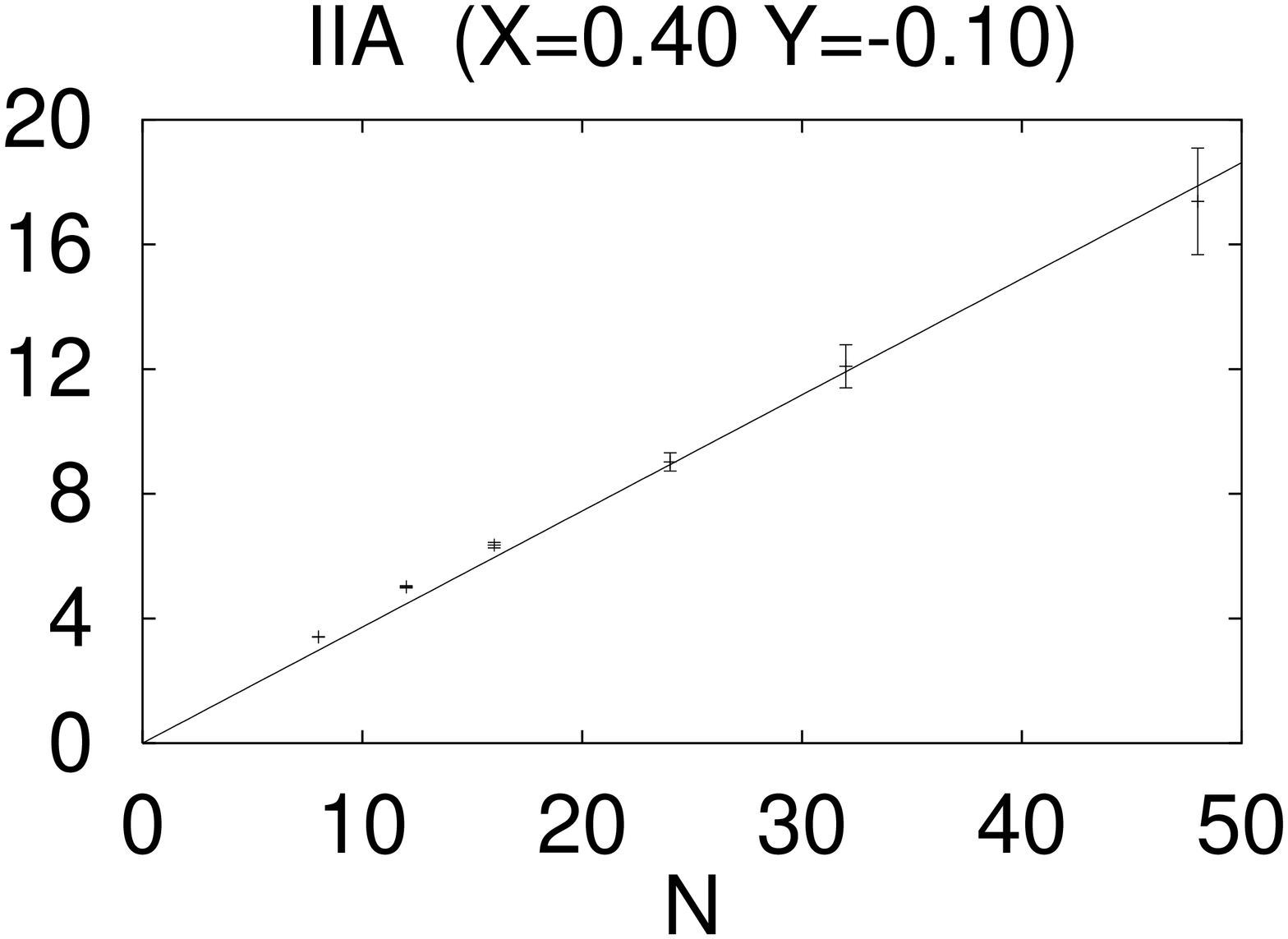}}} & 
  \scalebox{\hscale}[\vscale]{\rotatebox{-90.0}{\includegraphics{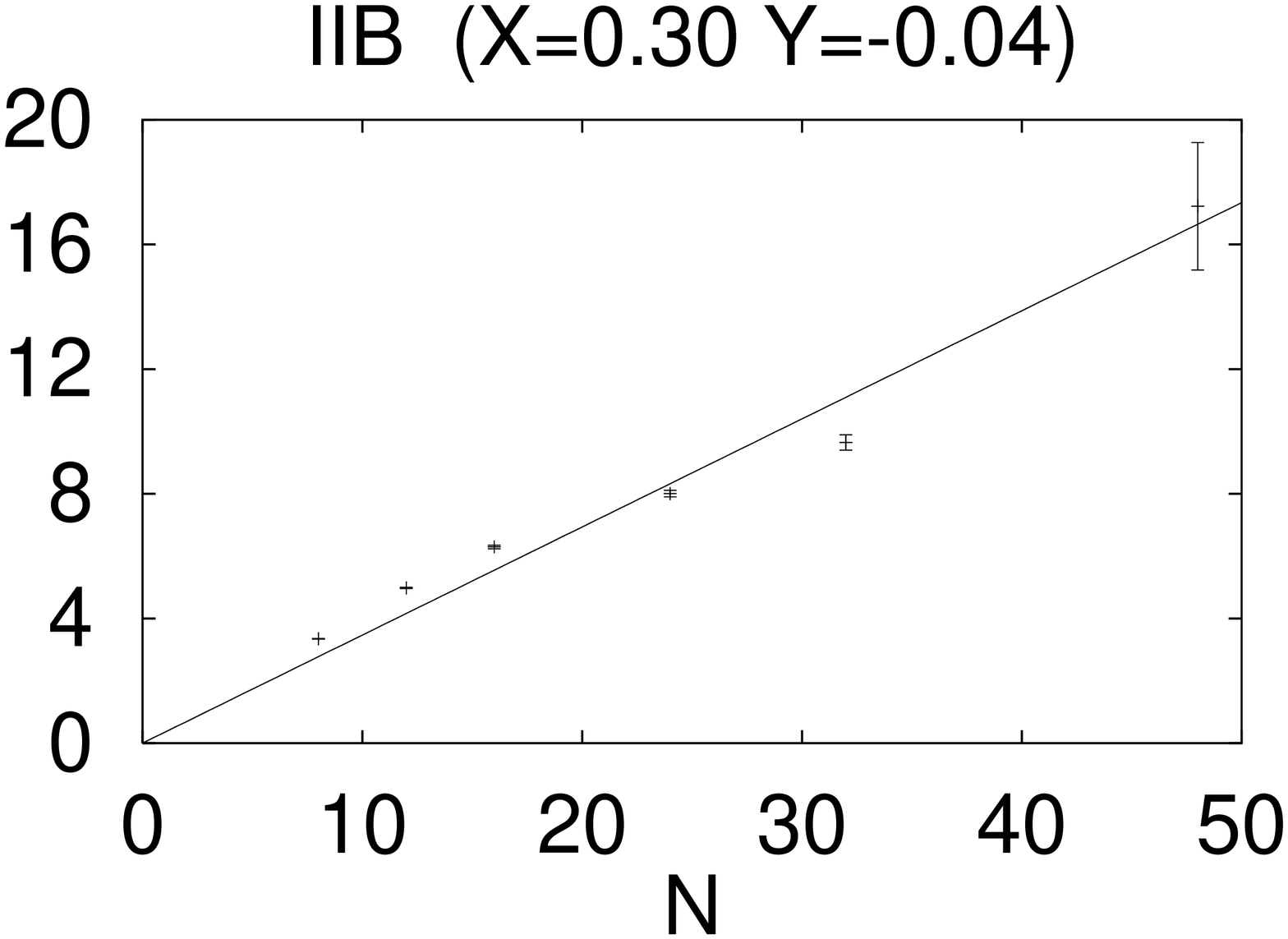}}} 
\end{tabular}  
\end{center}
\caption{\label{FSS}The ensemble averaged separation $\langle d \rangle$ between
  the two vortices, as a function of system size. The two upper
  figures are indicative of  \typeI behavior, whereas the two lower ones 
  indicate \typeII behavior.}
\end{figure}

In our simulations, the value of $n$ has been fixed to $n = 2$.
This corresponds to the case of  two field-induced vortices  which  move 
around in the system under the influence of their mutual effective 
interaction. During the simulations, we have measured the transverse position 
$r_{\perp}(z)$ of these two vortices labelled by $1$ and $2$, and the average 
distance between them.
\begin{equation}
  \label{VDist}
  \vdist = \frac{1}{N_z} \sum_{z} \left| r^1_{\perp}(z) - r^{2}_{\perp}(z)\right|.
\end{equation}
For \typeI superconductors this distance should be independent of system size, whereas 
for \typeII we expect that this distance scales with the system size. Finite size 
scaling of $\vdist$ for various points in the $(x,y)$ phase diagram is shown in
\Figref{FSS}.

In the part of phase diagram which we focus on, namely the region
defined by the dotted line in Fig.1 of \Citeref{Mo:2001}, the vortex lines
are generally directed and almost straight, well defined line objects.
This can be seen either by directly taking snapshot pictures of the
vortex-line configurations of the system, or by computing the
mean-square fluctuations around a straight-line configuration,
$\langle|r^i_{\perp}(z)-r^i_{\perp}(0)|^2\rangle$, for \emph{one}
vortex line. This is in contrast to the situation in the vicinity of
the {\it critical} part of $y_c(x)$ in Fig. 1 of  \Citeref{Mo:2001}, 
where the vortex lines loose their line tension via a vortex-loop blowout
\cite{Tesanovic:1999,Nguyen:1999}. Consequently, we can consider the
theory as an effective theory for an interacting pair of straight
vortex lines which interact with the dimensionless potential
$V(\vdist)$. If we make this assumption, the probability of finding
the vortices separated by a distance $\vdist$ in a system of 
$N^3$ lattice points with \emph{periodic boundary conditions}, 
is given by
\begin{equation}
  \label{PR}
  P_N(\vdist) = \frac{e^{-V(\vdist)} \Omega_N(\vdist)}{Z_{\vdist}},
\end{equation}
where $\Omega_N(\vdist)$ is the number of configurations with a transverse
vortex-vortex distance of $\vdist$, and $Z_{\vdist}$ is just a
normalisation factor. $\Omega_N(\vdist)$ can be calculated, either
analytically in the continuum limit
\begin{equation}
  \label{NRC}
  \Omega_N(\vdist) = 
  \begin{cases}
    2 \pi \vdist                                                         & \vdist < \frac{N}{2} \\
    2N\left(\frac{\pi}{2}-2 \arccos\left(\frac{N}{2\vdist}\right)\right) & \frac{N}{2} < \vdist < \frac{N}{\sqrt{2}}, 
  \end{cases}
\end{equation}
or by simple geometric counting in the case of a lattice. In the case
of noninteracting vortices, i.e. $V(\vdist) = 0$, the expectation
value of $\vdist$ is determined only by $\Omega_N(\vdist)$, and we find the
numerical value
\begin{equation}
  \label{dzero_def}
  \vdist_0 \equiv \langle \vdist \rangle = \frac{1}{Z_{\vdist}} 
\IntL{0}{\frac{N}{\sqrt{2}}} \d{\vdist} ~ ~ \Omega_N(\vdist) \vdist \approx 0.38 N.
\end{equation}

\begin{figure}[htbp]
\begin{center}
\begin{tabular}{cc}
  \scalebox{\hscale}[\vscale]{\rotatebox{-90.0}{\includegraphics{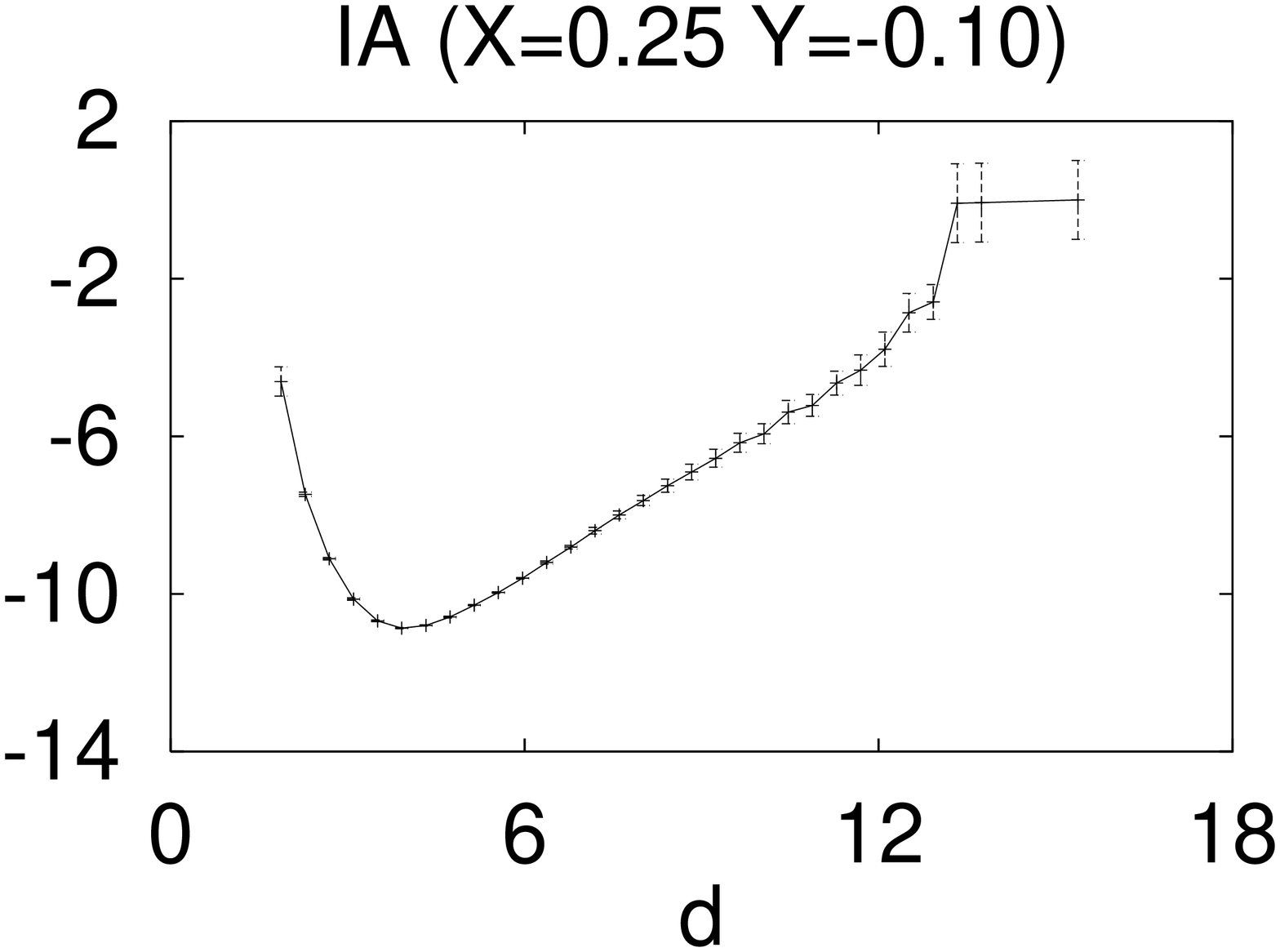}}} & 
  \scalebox{\hscale}[\vscale]{\rotatebox{-90.0}{\includegraphics{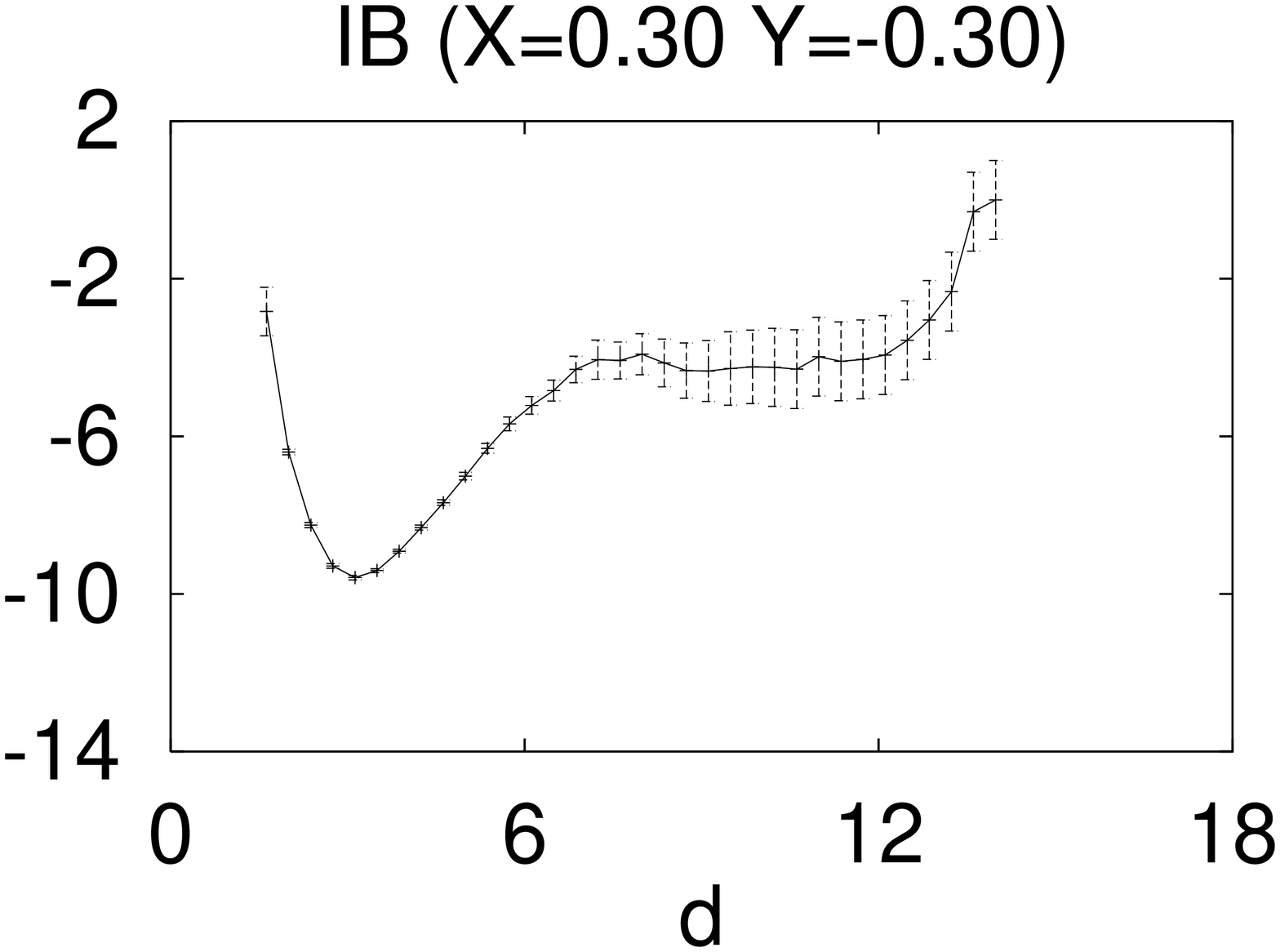}}}
\\                                                               
  \scalebox{\hscale}[\vscale]{\rotatebox{-90.0}{\includegraphics{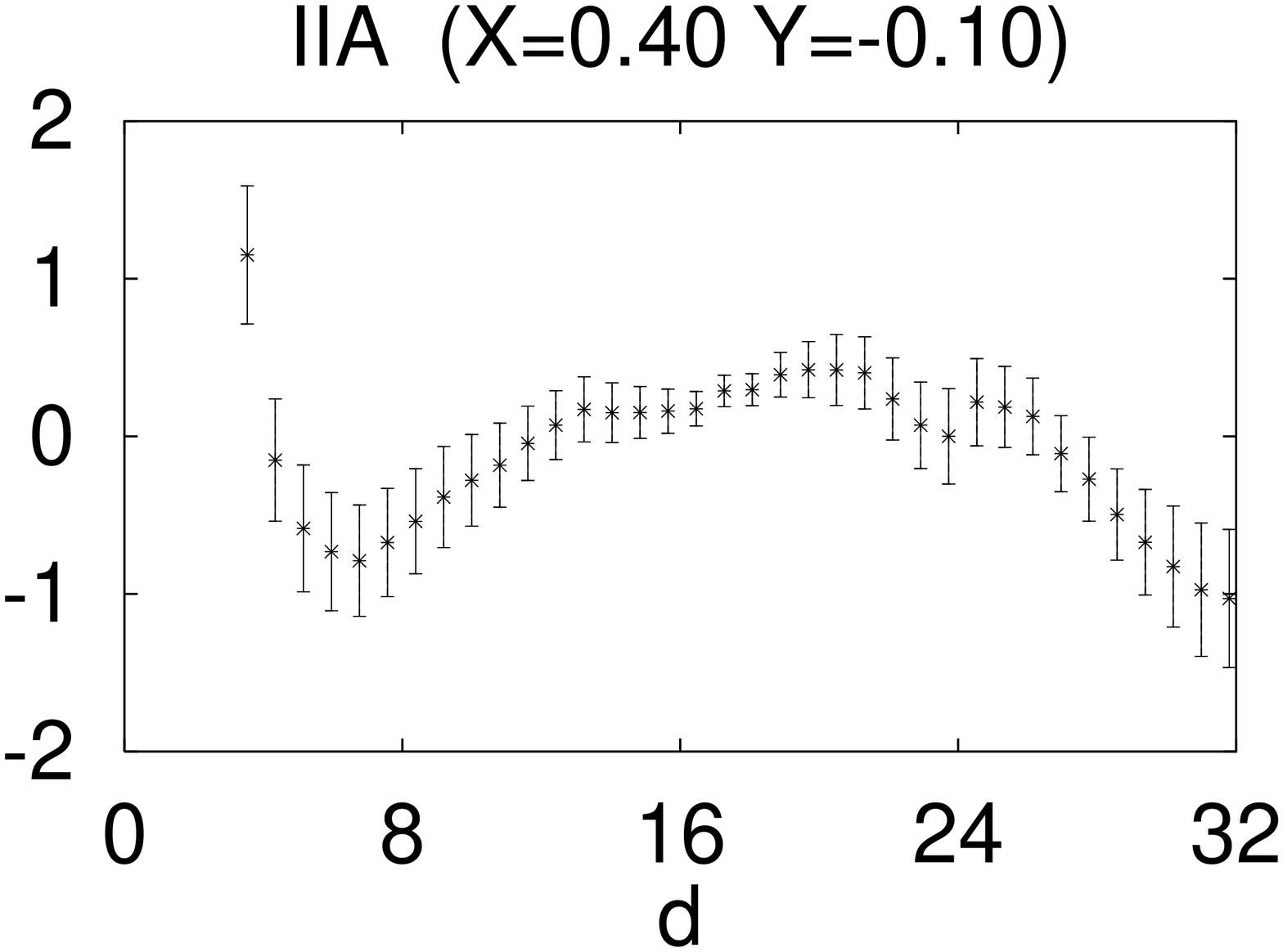}}} & 
  \scalebox{\hscale}[\vscale]{\rotatebox{-90.0}{\includegraphics{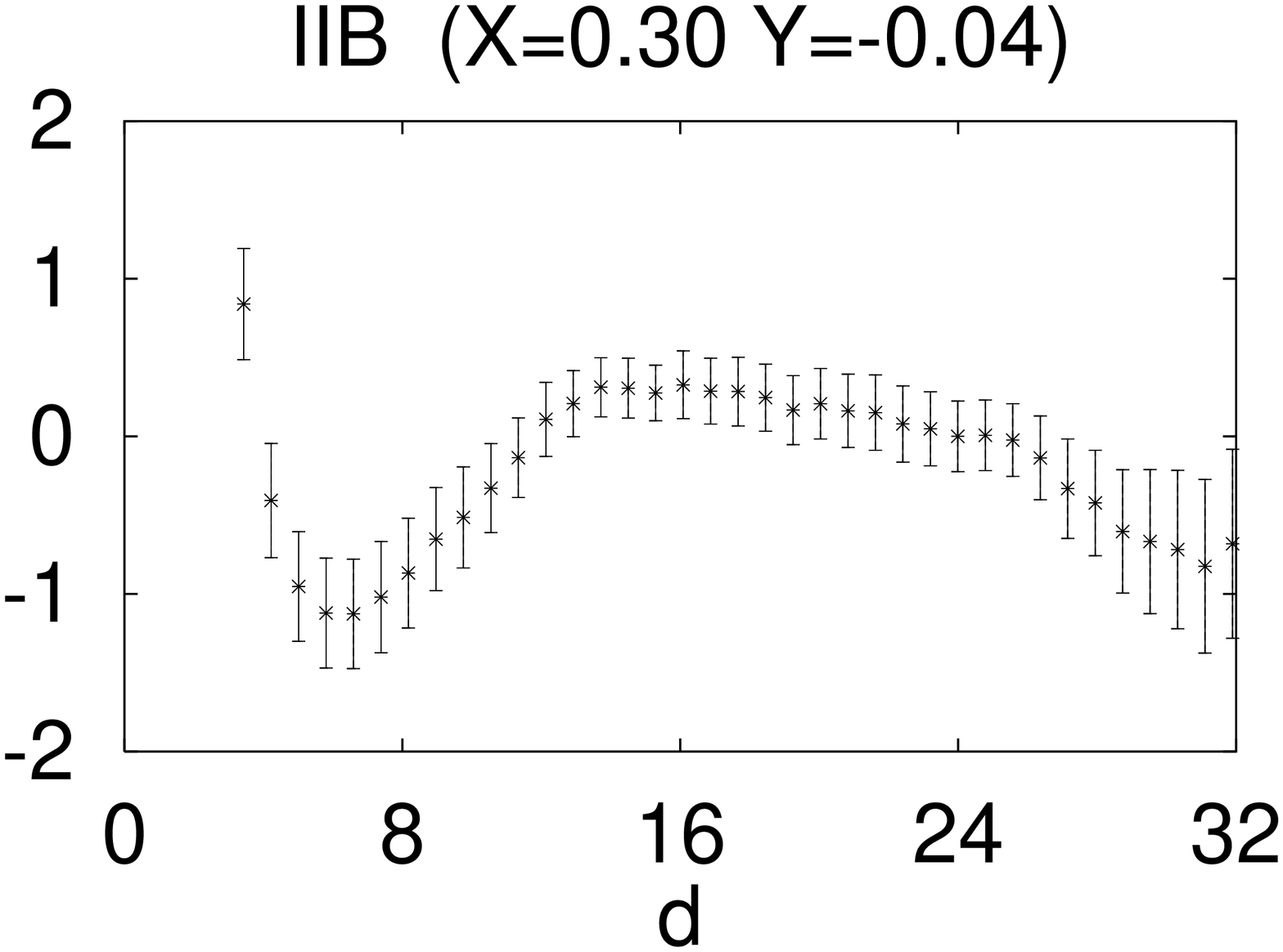}}} 
\end{tabular}  
 \end{center}
\caption{\label{VFig}The effective interaction  potential between vortices 
  $V(\vdist)$ as determined from \Eqref{PR}. Observe the difference in
  vertical scale, in the lower panels (\typeII) the interactions are
  much weaker than in the upper panels (\typeI). The graphs
  correspond to the same points in $(x,y)$ as those in \Figref{FSS}.}
\end{figure}

The separation $\vdist_0$ defined in \Eqref{dzero_def} will be used to
establish a numerical value of $\xIII$. Namely, we can compute the
averaged distance between vortices at fixed $x$ varying $y$, or vice
versa. In the latter case, we will use the criterion that if
$\langle\vdist\rangle$ exceeds som value $c \vdist_0$ where $c$ is
some fraction, then we have \typeII behavior, otherwise it is \typeI.
The quantity $\langle \vdist \rangle$ at fixed $y$ will turn out to be
an S-shaped curve as a function of $x$, increasing from small values
to large values as $x$ is increased. We interpret this as yet another
manifestation of the crossover from \typeI to \typeII behavior, and we
have chosen to locate the crossover region $\xIII$ at the value of $x$
where the curves change most rapidly, which is roughly when $\langle
\vdist \rangle \approx \vdist_0/2$.  As we shall see (see
\Figref{newFaseDiagram}), different crossover criteria give consistent
results.  The quantity $P_N(\vdist)$ can be estimated from histograms,
and then we can use \Eqref{PR} to determine the pair potential.
Depending on whether we consider \typeI or \typeII superconductors we
expect to see an attractive or a repulsive potential. \Figref{VFig}
shows the potential $V(\vdist)$ for the same points of the phase
diagram as \Figref{FSS}.

\subsection{Free energy}
\label{section:free.energy}
In \Eqref{fieldGL} we have used $n$ to indicate an integer number of
flux tubes, but in principle there is no reason to limit $n$ to
integer values, and we will use $S[\alpha,\psi,m]$ to denote a
generalisation to real $n$.

We have considered the free energy difference between a state
containing zero vortices, i.e. $m=0$ and a state containing $n$
vortices. We can not measure absolute values of the free energy, but
by differentiating\cite{Kajantie:1999}
\begin{equation}
  \label{free.energy}
  e^{-F(m)} = \Tr e^{-S(m)}
\end{equation}
with respect to $m$, and then integrating
up to $n$, we can calculate $\Delta F(n) = F(n) - F(0)$,
\begin{equation}
  \label{deltaF}
  \begin{split}
    &\frac{\Delta F(n)}{L_z \charge ^2} = \\
    &2\pi \beta_G \Int_0^{n} \d{m} \bigg[\underbrace{2m + \frac{1}{\pi N_z} 
     \left\langle\sum_{z} \alpha_{12}(x_0,y_0,z) \right\rangle_{m}}_{\equiv W(m)} \bigg].
  \end{split}
\end{equation} 

\begin{figure}[htbp]
\centerline{\scalebox{0.35}{\rotatebox{-90.0}{\includegraphics{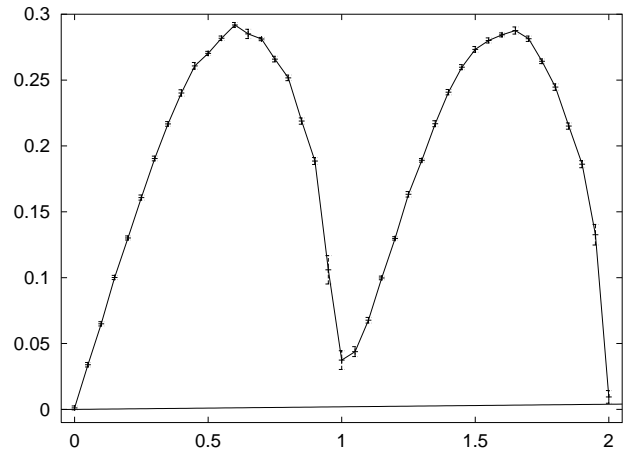}}}}
\caption{\label{wmfig} $W(m)$ The straight line corresponds to
  $2m/N_xN_y$ which according to \Eqref{wm:hel.tall} should be
  satisfied for $m$ integer.}
\end{figure}

To calculate $\Delta F$, we have then varied $m$ in steps of $\Delta m
= 0.05$, and performed the integration in \Eqref{deltaF} numerically.
Using shift symmetries, it can be shown \cite{Kajantie:1999} that $W(n)$
is equal to
\begin{equation}
  \label{wm:hel.tall}
  W(n) = \frac{2n}{N_xN_y},
\end{equation}
and the behavior for intermediate real values is shown in
\Figref{wmfig}. Increasing $n$ from $0$ to $1$ costs a free energy
$\Delta F(1)$, and adding two vortices costs an amount $\Delta F(2)$.
We will  \emph{always} have $\Delta F(2) > \Delta F(1)$, but the
question is whether $\Delta F(2) \gtrless 2 \Delta F(1)$. 
We may regard $F(n+2)+F(n)-2F(n+1)$ as the discrete second derivative
of the free energy with respect to particle number, which is nothing
but the inverse compressibility $K^{-1}$ of the vortex-system. 
In the thermodynamic limit this quantity can never become negative. 
However, its vanishing signals the onset of {\emph{phase-separation}} 
of the vortex system, which we again interpret as a lack of stability 
of the vortex-loop tangle, characteristic of \typeI behavior.

\subsection{Vortex compressibility, separation, and crossover}
\label{section:line.tension}
\begin{figure}[htbp]
\centerline{\scalebox{0.35}{\rotatebox{-90.0}{\includegraphics{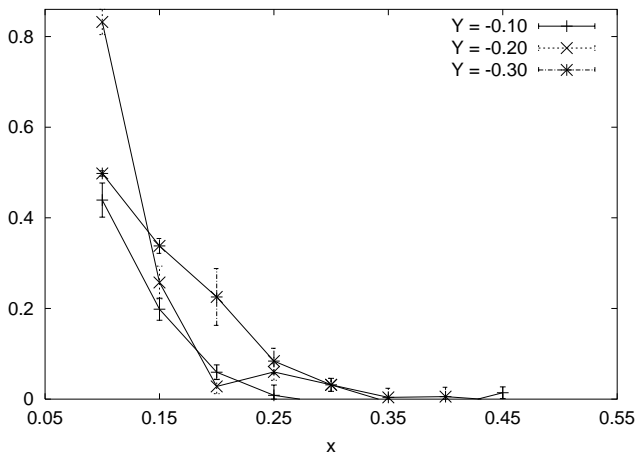}}}}
\caption{\label{deltaTFig} $\deltaT(x)$ for different values of $y$. The
  attraction vanishes for $\xIII < 0.5$, and $\xIII(y)$ is an
  increasing function of $|y|$.}
\end{figure}

We next define a quantity $\Delta T$ by the relation 
\begin{equation}
  \label{deltaT}
  \Delta T = \frac{\Delta F(1)}{L_z} - \frac{\Delta F(2)}{2L_z},
\end{equation}
which means that $\Delta T$ measures the relative free-energy difference between 
adding one vortex to the system  and half of that adding two vortices
to the system. Intuitively it is therefore clear that it  
measures the sign of the vortex interactions, and hence determines 
whether we are in the \typeI or \typeII regime. $\deltaT > 0$ signals  
attractive interactions, i.e. \typeI behavior, whereas $\deltaT \le 0$ 
signals repulsive interactions, i.e. \typeII behavior. We have calculated 
$\deltaT$ by using \DoubleEqref{deltaF}{deltaT}, the results are shown in
\Figref{deltaTFig}. The main qualitative result from these simulations 
is again that $\xIII(y)$ is a declining function of $y$. Note also
that $K^{-1} = - \Delta T$, and hence a positive $\Delta T$ clearly implies
phase-separation and instability of the vortex system, characteristic of
\typeI behavior. This is precisely what we see  for small $x$ in 
\Figref{deltaTFig}.

\begin{figure}[htbp]
\centerline{\scalebox{0.35}{\rotatebox{-90.0}{\includegraphics{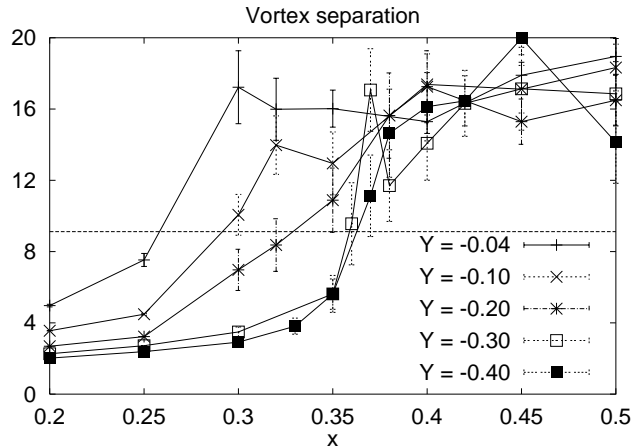}}}}
\caption{\label{DistFig} The ensemble averaged distance between a pair of 
  vortices, $\langle \vdist \rangle$, as a function of the square of
  the Ginzburg-Landau parameter $x = \kappa^2$, for various values of
  the temperature-like variable $y$. The horizontal line is at
  $\vdist_0 /2$.  Increasing $y$ amounts to increasing the
  temperature.}
\end{figure}

Finite size scaling of $\langle \vdist \rangle$ and studies of
$\deltaT$ differentiate nicely between strongly \typeI and \typeII
superconductors, but it is difficult to locate a value of $\xIII(y)$
with any great precision. \Figref{DistFig} shows $\langle \vdist \rangle (x)$ for
different values of $y$, along with a horizontal line at $\vdist_0/2$,
where $\vdist_0$ is the average separation between vortices had they been
non-interacting. We have found that $\vdist_0 \approx 0.38 N$ in our simulations.
We have, rather arbitrarily, taken the interception of this horizontal
line with the curve $\langle \vdist \rangle(x)$ as $\xIII$. 

\begin{figure}[htbp]
\centerline{\scalebox{0.24}{\rotatebox{0.0}{\includegraphics{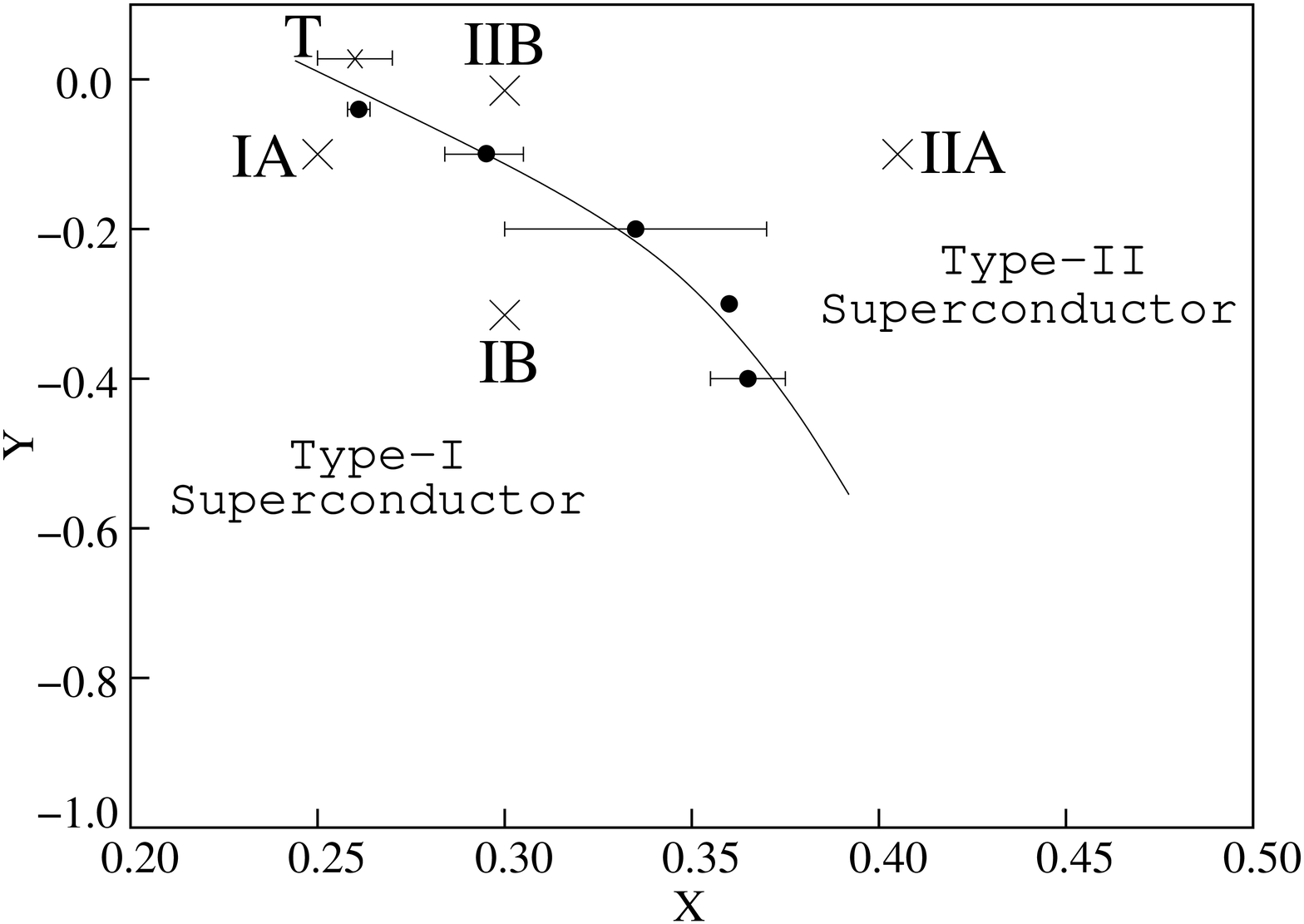}}}}
\caption{\label{newFaseDiagram} The computed crossover line  
  $y_{\rm{I/II}}(x)$ separating \typeI and \typeII superconductivity.
  The black circular filled points are given by the intercepts between
  $\vdist_0/2$ and the curves in \Figref{DistFig}. The four points
  labelled by (\IA, \IB, \IIA, \IIB) are the ones that were considered
  in detail in \DoubleFigref{FSS}{VFig}. The point marked \emph{T} is
  the \emph{tricritical} point in Fig. 1 of \Citeref{Mo:2001}, see the
  discussion of finite size effects in \secref{discussion.simulation}.
  The solid line connecting the points is \emph{purely} a guide to the
  eye. The computed line above corresponds to the dotted  line of Fig. 1 
  in  \Citeref{Mo:2001} in the vicinity of $(x_{\rm{tri}},y_{\rm{tri}})$.}
\end{figure}

The curves of $\langle \vdist \rangle(x)$ do not get significantly 
sharper with increasing system size, and there are no particular 
sharp features in $S[\alpha,\psi,2]$ as $x$ is increased  beyond $\xIII$. 
\Figref{newFaseDiagram} shows the intercepts from \Figref{DistFig}.
Due to the features in the curves of \Figref{DistFig}, and how the results
of \Figref{newFaseDiagram} are obtained from them, we tentatively
conclude that the computed line of \Figref{newFaseDiagram}, corresponding to 
the dashed line of Fig. 1  in  \Citeref{Mo:2001}, is a \emph{crossover} 
and \emph{not} a phase transition. However, we comment further on this in 
the concluding section.

As already indicated, there is some arbitrariness in the location of
$\xIII(y)$ in \Figref{newFaseDiagram}, however the four points
labelled by (\IA, \IB) and (\IIA, \IIB) clearly are in the \typeI and
\typeII regimes, respectivly. This is demonstrated in
\DoubleFigref{FSS}{VFig}.

\section{Discussion} 
From \DoubleFigref{FSS}{VFig} we conclude that there is a crossover
line separating effective attractive vortex interactions from
effective repulsive ones, i.e. \typeI and \typeII. This line can
either be crossed by changing $x$, i.e. \IA $\to$ \IIA, or by changing
the temperature i.e.  \IB $\to$ \IIB in \Figref{newFaseDiagram}. 
This means that for $x$ values in a suitable range, we can 
have in principle have a \emph{temperature induced} crossover from \typeI to \typeII 
superconductivity. Finally, we note that $\xIII(y)$ deviates significantly from the
mean-field value of $\xIII = 0.5$.

Deep in the \typeI regime, we find clear evidence of \emph{attractive}
interactions. In the \typeII regime the repulsive interactions appear
to be weak, and the results are essentially also consistent with two
\emph{randomly placed} vortex lines, i.e. not interacting, but not
consistent with an attractive force between the vortices. Therefore,
what the results unequivocally show is that by fixing the material
parameter $x$ and varying the temperature-like variable 
$y$, the character of the effective pair-potential is altered
inside the superconducting regime, significantly away from the critical 
line.

\subsection{Simulations}
\label{discussion.simulation}
The simulations with $n=2$, and the simulations with a real $m \in
[0,2]$ are quite different, and we will discuss them in
turn. An important feature of  \emph{all the simulations} 
in the present work is \emph{slow dynamics}.

For the $n=2$ simulation, where we have monitored $\vdist$, we find
that deep in the \typeI regime the simulations are quite
straightforward, the vortices stay close together with only small
fluctuations, and a moderate number of sweeps is sufficient to get good
statistics. However when we increase $x$ towards $\xIII$ the effect is
\emph{not} that $\vdist$ stabilises at a higher value, instead we get
get fluctuations between a \typeI like state where the two vortices
are close together, and a \typeII like state with large vortex-vortex
separation. This picture persists as $x$ is increased into the \typeII
regime, the only difference is that the fraction of time spent in the
\typeI like state state decreases.  In fact the results of
\Figref{FSS} and \Figref{VFig} are in the \typeII region quite close
to what we would get from noninteracting vortices.  What makes the
simulations difficult is that moving a vortex line in the transverse
direction is a \emph{global} change, and thereby very slow. Timeseries
of $\vdist$ show characteristic time scales of $10^4$ sweeps, so 
long simulations $\sim 10^6$ sweeps over the lattice are required 
to obtain acceptable accuracy. A truly high-precision determination 
of $\xIII(y)$ would surely benefit from a specialized algorithm for the 
MC updates.

To get good results one should take the $N\to \infty$ limit. The
conclusions from the results of  \Figref{FSS} are based on this limit, 
whereas those drawn from Figs.
\ref{VFig},\ref{DistFig} and \ref{newFaseDiagram} are based on the
fixed system size $N = 48$. We have not performed a systematic study
of finite size effects, but the curves in \Figref{DistFig} \emph{do have}
subject to finite size effects in them. The trend is that curves move to
the right upon increasing system size, this very likely explains the
apparent discrepancy between the tricritical point (where the $N \to
\infty$ limit has been applied), and the remaining points in
\Figref{newFaseDiagram}. 

Note that \Eqref{GLM} is a continuum field theory,
and $\xIII(y)$ is \emph{not} a critical point, hence the continuum
limit $\beta_G \to \infty$ should be taken. Our experience
from the large-scale simulations performed in \Citeref{Mo:2001} 
indicates that $\beta_G = 1.00$ provides
conditions in the simulations already quite close to the continuum 
limit. We have therefore chosen to work with $\beta_G=1.00$
and focused our efforts on considering  large systems and long 
simulations.

The Monte-Carlo computations of  $\deltaT$ have been even more time consuming,
because we have had to do the simulations for 41 different values of $m$. 
We have therefore limited ourselves to considering only the system $L=24,
\beta_G = 1.00$, for a discussion of finite $N$ / finite $\beta_G$
effects see \Citeref{Kajantie:1999}. The relaxation time for these
simulations has been particularly long in the limits $m \to 1^{-}$ and $m \to
2^{-1}$, and we therefore have performed much longer simulations in these
limits than for intermediate values of $m$.
From \Figref{wmfig} and \Eqref{wm:hel.tall} it is seen that $W(m)$ is
a quantity of order $\mathcal{O}(N^{-2})$ whose finiteness originates
in the difference between two $\mathcal{O}(1)$ quantities.
Consequently, it is difficult to get numerically precise results. This
has in particular been the case in the limits $m \to1^{-}$ and $m \to
2^{-}$.

\subsection{Crossover}
The physical picture that emerges for the thermal renormalization of the
vortex interactions, is the following. At low temperatures, for the $\kappa$-values
we consider, the system is in the \typeI region with a fairly deep minimum
in the effective pair-potential between vortices at short distances, leading to
an attractive interaction. At very short distances, we find a steric repulsion
on the scale of the lattice constants in the system due to the  large
Coulomb barrier that must be overcome to occupy a link with
two or more elementary vortex segments. This
lenght scale represents the size of the vortex core in the problem.
Upon increasing the temperature to the vicinity of the line $y_c(x)$,
we do not find large transverse meanderings of the individual 
vortex lines as we move along each vortex line, rather the vortex lines
are essentially straight. Therefore, we believe that it is {\it not} 
entropic repulsion due to the bare steric repulsion in the problem,
of the type which it seems reasonable to invoke for {\it strongly} 
fluctuating elastic strings \cite{Zaanen:2000,Mukhin:2001} that 
renormalizes the vortex interactions in the way that is seen in \Figref{VFig}.  
Rather, what appears to happen  is that the vortex lines slosh back and forth 
in the minimum of the effective potential well as essentially straight lines. 
Hence,  to a larger and larger extent as temperature is increased, 
they experience the hard wall in the interactions at small distances, 
and the weak attraction at large distances, effectively washing out 
the minimum in the potential, thus making it effectively  more repulsive.  

This is also seen in our simulations (not shown in any of the figures)
when we monitor the transverse meandering fluctuations of each vortex
line, $\langle|r_{\perp}(z)-r_{\perp}(0)|^2\rangle$, as well as the
mean square fluctuations of the intervortex distance,
$\langle\vdist^2\rangle-\langle\vdist\rangle^2$, where $\vdist$ is
defined in Eq. (\ref{VDist}). The former is small deep in the
superconducting regime, and remains small as the line $y_c(x)$ in
Fig. 1 of \Citeref{Mo:2001} is approached, while the latter increases
dramatically as the dotted crossover line is crossed. It is precisely
this fact which makes the simulations extremely time consuming.

One should however keep in mind that, since we are considering the 
full GL theory in our simulations, and not the linearized London
limit,  it is in any case a drastic simplification to view the effective 
interaction between vortices as a simple pair potential.

Finally, we note that, although our present simulations, which by 
necessity are on finite-sized systems, indicate that the 
change from \typeI to \typeII behavior is a crossover, we cannot
rule out the possibility that it is elevated to a true 
phase-transition in the thermodynamic limit. More work is
needed to clarify if this is indeed the case, but this will 
have to await the next generation of massive parallell 
computers. Questions that need to be addressed in this context,
are: What is the order parameter of such a transition, and  
what symmetry, if any, is being broken. 

\section{Conclusion} 
We have considered the effective interaction between
two vortices in the full GL model, and how this effective
interaction is influenced by thermal fluctuations. We have included 
fluctuations in the gauge fields, as well as the phase- and 
amplitude-fluctuations of the complex scalar matter field of the 
problem. We have found that the effective interaction changes from
being attractive to being repulsive  at $\xIII$. This means a change from
\typeI to \typeII behavior. We have found that  $\xIII$ is below
the standard quoted value of $0.5$, and is a function of the temperature-like
parameter $y$. This means that at the critical point, the value of the 
GL parameter that separates \typeI from \typeI behavior is 
smaller than $1/\sqrt{2}$. The line $\xIII(y)$ appears  to be a crossover, 
and not a true phase transition. The above seems to offer a simple explanation
for the experimental observation that elemental \metal{Ta} and \metal{Nb} 
superconductors show a crossover from \typeI to \typeII behavior as the 
temperature is increased towards $T_c$. Previous explanations based on 
mean-field theories and not involving thermal fluctuations required two
additional temperature dependent $\kappa$-values to be defined \cite{Maki:1969}.

We acknowledge support from the Norwegian Research Council via
the High Performance Computing Program (S.M.,J.H.,A.S.), and Grant Nos. 
124106/410 (S.M.,A.S.) and 148825/432 (A.S.), and A.S. thanks E. H. Brandt 
for useful comments. J.H. also acknowledges support from NTNU via a university 
fellowship.

\end{document}